\documentstyle[12pt,epsfig]{article}
\begin{document}
\setlength{\parskip}{0.5cm}
\setlength{\baselineskip}{0.65cm}
\begin{titlepage}
\begin{flushright}
CCL-TR-95-004 \\ DESY 95-096 \\ May 1995
\end{flushright}
\vspace{0.5cm}
\begin{center}
\Large
{\bf Constraints on the Proton's Gluon Distribution} \\
\vspace{0.1cm}
{\bf from Prompt Photon Production} \\
\vspace{1.5cm}
\large
W. Vogelsang \\
\vspace{0.5cm}
\normalsize
Rutherford Appleton Laboratory \\
\vspace{0.1cm}
Chilton Didcot, Oxon OX11 0QX, England \\
\vspace{1.2cm}
\large
A. Vogt$\, ^{\ast}$\\
\vspace{0.5cm}
\normalsize
Deutsches Elektronen-Synchrotron DESY \\
\vspace{0.1cm}
Notkestra{\ss}e 85, D-22603 Hamburg, Germany \\
\vspace{2.5cm}
{\bf Abstract}
\vspace{-0.3cm}
\end{center}
We analyze the capability of prompt photon production in $pp$ and
$p\bar{p}$ collisions to constrain the gluon distribution of the
proton, considering data from fixed-target experiments as well as
collider measurements. Combined fits are performed to these large-$p_T$
direct-$\gamma $ cross sections and lepton-proton deep-inelastic
scattering data in the framework of next-to-leading order perturbative
QCD. Special attention is paid to theoretical uncertainties originating
from the scale dependence of the results and from the fragmentation
contribution to the prompt photon cross section.
\vfill
\noindent
\small
$^{\ast} $ On leave of absence from Sektion Physik, Universit\"at
M\"unchen, D-80333 Munich, Germany
\normalsize
\end{titlepage}
\section{Introduction}
The production of high-$p_T$ prompt photons in $pp$ or $p\bar{p}$
collisions provides an important probe of the proton's gluon
distribution, $g(x,Q^2)$, due to the presence and dominance of the
leading order (LO) ${\cal O}(\alpha \alpha_s)$ `Compton-like' subprocess
$ qg\rightarrow \gamma q $. In fact, constraints on $g(x,Q^2)$ for
$ 0.3 \stackrel{<}{{\scriptstyle \sim}} x \stackrel{<}{{\scriptstyle
\sim}} 0.6$ at $Q^2 \stackrel{<}{{\scriptstyle \sim}} 10 \mbox{ GeV}^2
$, derived mainly from the WA70 fixed-target $pp \rightarrow \gamma X $
data \cite{wa70}, have been the backbone of the gluon determination
in many parton density analyses \cite{mrs,grvold,grv,cteq} ever since
the pioneering work of \cite{aur1}. In that paper, a combined
next-to-leading order (NLO) fit to the WA70 data and to deep-inelastic
scattering (DIS) results from the BCDMS collaboration \cite{bcdms}
was performed, and available direct photon data from other fixed-target
experiments \cite{na24,ua6old} as well as from ISR \cite{r110,r806} and
$Sp\bar{p}S$ \cite{ua1,ua2old} were compared to this fit. Since then,
major theoretical and experimental developments concerning
direct-$\gamma $ production have taken place, and HERA results begin to
add DIS constraints in a previously unexplored kinematic region.
Therefore, we feel that it is time now for a reanalysis of the prompt
photon data and their implications on the gluon density.

Experimentally, much progress has been made during the last years. Some
of the direct photon data sets mentioned above have been superseded by
improved analyses \cite{r807,ua2,ua6}, usually providing smaller
statistical and systematic errors. Even more importantly, the accessible
range of fractional gluon momenta $x$ has been considerably enlarged by
the recent, partly very precise measurements at the Fermilab Tevatron
\cite{cdfalt,cdf,d0}. By now, the region $ 0.01 \stackrel{<}
{{\scriptstyle \sim}} x \stackrel{<}{{\scriptstyle \sim}} 0.6 $ is
completely covered, with the data from WA70 \cite{wa70}, R806
\cite{r807}, UA2 \cite{ua2} and CDF \cite{cdf} presently dominating in
their respective kinematical regimes.
Also, the DIS kinematic coverage has been dramatically improved by the
NMC \cite{nmc} and recent HERA \cite{zeusf2,h1f2} $ F_2^{\, p}(x,Q^2) $
structure function data. Quark density measurements, extending down to
almost $ x = 10^{-4} $ now, also imply an important constraint on the
gluon distribution due to the momentum sum rule. Moreover, scaling
violations of $ F_2^{\, p} $ at HERA now begin to constrain $g(x,Q^2)$
severely, especially at $ x \stackrel{<} {{\scriptstyle \sim}} 0.01 $
\cite{svzeus}. Hence it is interesting to examine quantitatively the
questions of whether a successful NLO perturbative QCD description of
all these data is possible and how much freedom is left for $g(x,Q^2)$.

On the theoretical side, it was not yet possible to perform a complete
and fully consistent calculation of the NLO prompt photon cross section
at the time of the analysis in \cite{aur1}, since the NLO fragmentation
contribution, based on the partonic $2 \rightarrow 3$ QCD subprocesses
\cite{acgg} and on corresponding parton-to-photon fragmentation
functions \cite{aurf,grvf}, was not yet available. Also, the development
of a proper NLO theoretical implementation of isolation cuts
\cite{bq,boo}, imposed on the cross section in the high-energy
$Sp\bar{p}S$ and Tevatron experiments, was only recently finished
\cite{gv1} and demonstrated to be phenomenologically important
\cite{ggrv}. Even in a very recent global study on direct-$\gamma $
production by members of the CTEQ group \cite{cteqdg}, the fragmentation
pieces are included only in LO and the isolated collider data are
transformed to fully inclusive cross sections on this basis.

Finally, the theoretical uncertainties still present in the NLO
treatment have not been considered to their full extent in the
literature so far. For instance, the dependence of the theoretical
cross section on unphysical scales, such as the renormalization scale
$\mu_R$ and the factorization scale $\mu_F$, has been treated in the
fits of \cite{aur1,mrsl} by optimizing the scales using the `principle
of minimal sensitivity' \cite{optim,pms}, i.e., by choosing scales where
the NLO cross section is stationary with respect to small changes in
$\mu_R$ and $\mu_F$. While this concept certainly is attractive to
some extent, see however \cite{bq} for a critical discussion, it
inevitably does suppress the uncertainty of the theoretical prediction
due to the scale dependences. The recent study of \cite{cteqdg} has
not thoroughly addressed this issue either: in that analysis the scales
were included, subject to the constraint $\mu_R = \mu_F$, among the
parameters fitted to data, which represents just another kind of
optimization.  There are also other ambiguities, e.g.\ originating
from the experimentally virtually unknown parton-to-photon fragmentation
functions. In order to examine the question of how effective the
constraints on the gluon distribution coming from prompt photon
production really are, it is a crucial issue to take into account such
uncertainties inherent to the calculation.

The remainder of this paper is organized as follows: In section 2 we
briefly recall the main ingredients needed to calculate the inclusive
and isolated prompt photon cross sections in NLO. Section 3 is devoted
to a detailed discussion of the main theoretical uncertainties mentioned
above and their effects on the calculated cross sections. In section 4
we present combined analyses of an exhaustive set of DIS and prompt
photon data in order to arrive at conclusions about the quantitative
viability of the NLO framework and about the constraints on the gluon
distribution. Finally, we summarize our main findings in section 5.
\section{General Framework}
Two types of processes contribute to the prompt photon production cross
section: the so-called `direct' piece, where the photon is emitted via
a pointlike (direct) coupling to a quark, and the fragmentation piece,
in which the photon originates from the fragmentation of a final state
parton. Despite the fact that its corresponding partonic subprocesses
are of order $\alpha_s^2$, the fragmentation contribution is present
already in LO since the parton-to-photon fragmentation functions are
effectively of order $\alpha/\alpha_s$ in perturbative QCD, where
$\alpha$ denotes the fine structure constant.
Next order corrections in the strong coupling constant $\alpha_s$ have
been calculated in the $\overline{\mbox{MS}}$ renormalization and
factorization schemes for both the direct \cite{boo,optim,gv} and the
fragmentation \cite{acgg} subprocesses, hence the cross sections can
be consistently calculated to order $\alpha \alpha_s^2$. The cross
section for the fully inclusive production of a prompt photon with
momentum $p_{\gamma}$ schematically reads
\begin{eqnarray}
\lefteqn{d\sigma  \equiv d\sigma_{dir}+d\sigma_{frag}
=\sum_{a,b=q,\bar{q},g}\int dx_a dx_b
f_a(x_a,\mu_F^2) f_b(x_b,\mu_F^2) \; \times } \\
& & \left[ d\hat{\sigma}_{ab}^{\gamma}(p_{\gamma}, x_a,x_b, \mu_R,
\mu_F,M_F)+\! \sum_{c=q,\bar{q},g}\int^1_{z_{min}}\! \frac{dz}{z^2}
d\hat{\sigma}_{ab}^c (p_{\gamma},x_a,x_b,z,\mu_R,\mu_F,M_F)
D^{\gamma}_c (z,M_F^2)\right] \nonumber
\end{eqnarray}
where $ z_{min}=x_T \cosh \eta $ with the prompt photon's rapidity
$\eta$, and $x_T = 2p_T/\sqrt{s}$. In eq. (1), $d\hat{\sigma}_{ab}^i$
represent the subprocess cross sections for partons $a,b$ producing a
particle $i$ ($i=\gamma$, $q$, $g$), integrated over the full phase
space of all other final state particles. $f_i(x,\mu_F^2)$ denotes the
number density of the parton type $i$ in the proton (or antiproton) at
momentum fraction $x$ and scale $\mu_F $, and $D_c^{\gamma}(z,M_F^2)$
is the photon fragmentation function at scale $M_F$, $z$ being the
fraction of energy of the fragmenting parton $c$ transferred to the
photon.

As already mentioned in the introduction and made explicit in eq.\ (1),
the cross section in any fixed order of perturbation theory depends on
unphysical scales which have to be introduced in the procedure of
renormalization ($\mu_R$) and of factorization of initial ($\mu_F$)
and final ($M_F$) state mass singularities. The latter type of
singularities appears, e.g., in the calculation of the ${\cal O}
(\alpha_s^3)$ $ab\rightarrow cde$ NLO fragmentation subprocess
cross sections when the parton $c$ (which then fragments into the
photon) becomes collinear with particle $d$ or $e$, but also in the
calculation of the ${\cal O}(\alpha \alpha_s^2)$ $ab\rightarrow \gamma
de$ NLO `direct' subprocess cross sections when the photon and a final
state quark are collinear. These singularities need to be factorized at
a scale $M_F$ into the `bare' fragmentation functions in order to render
the cross section finite. The fragmentation functions then obey
corresponding NLO evolution equations.  Since factorizing singularities
is not a unique procedure but depends on the factorization prescription
adopted, it becomes obvious that only the sum of the direct and the
fragmentation pieces is a physical (scheme independent) quantity beyond
the LO, but not these parts individually. In particular, even if the
fragmentation contribution turns out to be numerically small, its
addition on a LO basis to a NLO `direct' piece is theoretically
inconsistent and yields a scheme dependent cross section. Needless to
say that a consistent NLO calculation also affords parton distributions
and photon fragmentation functions evolved according to their respective
NLO evolution equations. For calculating these $Q^2$-evolutions we use
the Mellin-$n$ space technique described in \cite{grv90}.

At very high-energy $p\bar{p}$ colliders the photon is experimentally
required to be `isolated' in order to suppress the huge background due
to copious production of $\pi^0$, the decay of which can fake a prompt
photon event. This means that the amount of hadronic energy $E_{had}$
allowed in a cone $\sqrt{(\Delta \phi)^2 + (\Delta \eta)^2} \leq R$
around the photon direction is limited to a small fraction of the photon
energy, $E_{had}\leq \epsilon E_{\gamma}$ with $\epsilon \stackrel{<}
{{\scriptstyle \sim}} 0.1$. In order to compare QCD predictions with
isolated collider data, the theoretical calculation has to include this
isolation criterion which leads to a significant decrease from the fully
inclusive cross section \cite{gv1,aurcoll}. In \cite{gv1} a simple, yet
accurate way of incorporating the isolation cut into the NLO calculation
has been developed. Starting from the fully inclusive cross section,
the isolated one is obtained by introducing a subtraction term
\cite{bq,gv1},
\begin{equation}
d\sigma^{isol}(R,\epsilon) =
d\sigma^{incl}-d\sigma^{sub}(R,\epsilon) \:\:\: ,
\end{equation}
where $d\sigma^{sub}(R,\epsilon)$ is the cross section for having
{\em more} hadronic energy than $\epsilon E_{\gamma}$ in the cone around
the photon. It turns out that $d\sigma^{sub}(R,\epsilon)$ can be easily
and reliably calculated in the approximation of a rather narrow cone. An
equation similar to eq.\ (2) (which holds for the NLO direct
contributions) can also be written down for the NLO fragmentation piece
\cite{gv1}, in which case the main effect of imposing the isolation cut
is to raise the lower integration limit in eq.\ (1) to $ z_{min} =
1/(1+\epsilon )$. In this way, it is possible to calculate also the
isolated prompt photon production cross section measured at high-energy
colliders in a consistent way beyond the leading order.

We conclude this section by noting that the recent study in
\cite{cteqdg} reports a (rather $p_T$ independent) ${\cal O}(10\%)$
discrepancy between the NLO programs of \cite{boo} and \cite{gv} in the
kinematic regime of the CDF measurements, which persists also if the
fully inclusive cross section is considered \cite{kuhl}. In this context
it is interesting to mention that there is no such discrepancy between
the two calculations of \cite {optim} and \cite{gv}, which are in {\em
exact} agreement concerning the NLO direct contribution to the fully
inclusive cross section for all $p_T$ and $\sqrt{s}$. For all
calculations to follow we use the program of \cite{gv} for the direct
part of the NLO prompt photon cross section, along with the expressions
of \cite{acgg} for the fragmentation contribution. When calculating the
isolated cross section we complement these programs according to the
prescription of \cite{gv1}.
\section{Theoretical Uncertainties}
In this section, we address the main uncertainties entering the NLO
calculation of the prompt photon cross section, namely the dependence on
the photon fragmentation functions and on the renormalization and mass
factorization scales. As a point of reference, we first calculate the
cross section for a fixed `standard' set of input distributions and
parameters and confront it with the data. For this purpose, the parton
distributions and photon fragmentation functions are taken from GRV
\cite{grv,grvf}, together with the value of the QCD scale parameter for
four active flavours, $\Lambda_{\overline{\rm MS}}^{(4)} = 200 $ MeV.
We choose $ \mu_R = \mu_F = M_F = p_T/2 $ for the renormalization and
factorization scales except for the isolated prompt photon data, where
$M_F=R p_T$ seems more appropriate \cite{bq}.

The very small charm effects in the cross section at fixed-target
and ISR energies are neglected. For the $Sp\bar{p}S$ and the Tevatron
experiments, however, charm-induced contributions coming, e.g., from
$cg\rightarrow \gamma c$ are not negligible. We employ the effective
(massless) charm quark distribution of \cite{grvold} in calculating
these contributions. We have checked that reasonable variations of
this charm density do not significantly alter our results. An
alternative approach to the heavy quark (charm) contribution is to
perform mass factorization only for the light $u$, $d$, and $s$ quarks.
In this scheme, also used in \cite{grv}, the heavy flavours $c$, $b,
\ldots $ do not act as partons in the proton, and in LO charmed prompt
photon events are only introduced via the processes $ gg\rightarrow
\gamma c \bar{c}$ and $q\bar{q}\rightarrow \gamma c\bar{c}$ with {\em
massive} charm quarks. The results of \cite{sv} show, however, that
these two approaches yield very similar results at least in LO. Thus the
theoretical uncertainty originating from the charm treatment seems to
be rather small. For the rest of this paper, we therefore use the
fixed intrinsic charm quark distribution of \cite{grvold} which
facilitates the calculations.

The sets of experimental data we take into account in this section are
the fixed-target data of WA70 \cite{wa70} ($\sqrt{s}=24$ GeV, $pp$
inclusive), the ISR results from R806/7 \cite{r807} ($\sqrt{s}=63$ GeV,
 $pp$ inclusive), the $Sp\bar{p}S$ results from UA2 \cite{ua2} ($\sqrt
{s} =630$ GeV, $p\bar{p}$ isolated) and the Tevatron data of CDF
\cite{cdf} ($\sqrt{s}=1.8$ TeV, $p\bar{p}$ isolated with $R=0.7$ and
$\epsilon=2$ GeV$/p_T$). As mentioned above, these data sets dominate
in their respective kinematical domains.  The cross sections of
\cite{wa70} and \cite{cdf} have been averaged over  the experimentally
covered regions of rapidity $\eta$, while the results in \cite{r807} and
\cite{ua2} have been presented at $ \eta = 0 $. For our comparisons and
fits, we add the statistical and systematic errors in quadrature, using
point-to-point errors where these are separately available \cite
{ua2,cdf}\footnote{We thank S. Kuhlman for providing the break-up of
the systematic errors of the CDF direct photon data into a
point-to-point and a fully correlated part.}.
Fig.\ 1 displays the results for our `standard' choice of input
distributions and parameters. We show the `default quantity'  (data
$-$ theory)/theory versus $x_T$. This provides a particularly easy
visualization of the (dis)agreement between data and theoretical
calculation in view of the strong $p_T$ fall-off of the cross section.
$x_T$ is a good representative of the Bj\o rken-$x$ values predominantly
probed in the gluon distribution at given $p_T$ and $\sqrt{s}$.

As can be seen from Fig.\ 1, the overall agreement between data and the
NLO theoretical prediction is good though not complete, especially if
the very small errors of the CDF data are taken at face value. The
agreement between theory and the fixed-target and ISR measurements is
very good, whereas the comparison with the high-energy collider results
seems to be slightly less successful, since both the CDF and the UA2
data show a somewhat stronger rise for small $x_T$ than the theoretical
cross section\footnote{Note, however, that there is good agreement
between NLO theory and the preliminary Tevatron D0 results \cite{d0},
which on the other hand have sizeably larger errors than the CDF data
\cite{cdf}.}.
This effect is more pronounced and statistically more meaningful for the
CDF results, which possess the smallest point-to-point errors of all
data sets and therefore provide a very precise measurement of the
{\em slope} of the cross section. Note that the CDF \cite{cdf} as well
as the UA2 \cite{ua2} data are subject to a normalization uncertainty
of about $10\%$ which we have used in the figure to center the results
on the zero line. It has to be emphasized that a {\em much} stronger
discrepancy between high-energy collider data and NLO calculations was
reported previously \cite{cdfalt,cdf}. As was shown in \cite{ggrv},
a dramatic improvement is obtained by using `modern' sets of (steep)
parton distribution functions like, e.g., those of GRV \cite{grvold,grv}
or the most recent MRS(A$^{\prime }$,G) sets \cite{mrs} as well as,
equally important, by including a properly isolated {\em NLO}
fragmentation contribution in the calculation. According to Fig.\ 1,
the latter amounts to a $20\%$ slope effect for CDF conditions, thus
its inclusion is clearly crucial for a quantitative comparison between
the  experimental and theoretical cross sections.
The fragmentation piece is non-negligible {\em despite} the presence
of the isolation cut without which it would easily contribute about
$50\%$ to the total cross section \cite{aurf,ggrv,aurcoll}. The results
in Fig.\ 1 indicate a (minor) remaining discrepancy between the
$p_T$-slopes of the experimental and theoretical cross sections even
after the improvements of \cite{ggrv} have been applied. One furthermore
infers from the figure that fragmentation also plays an important role
in the calculations in the ISR and fixed-target regions, where no
isolation cut has been applied. Here it leads to an effect of partly
even more than $20\%$, but influences the slope to a lesser extent.

An important uncertainty in the calculation is the dependence of
the cross section on the parton-to-photon fragmentation functions
which are experimentally unknown so far.
Two partly very different NLO sets of such distributions have been
suggested in the literature, namely in \cite{aurf} (ACFGP) and in \cite
{grvf} (GRV). Fig.\ 2 presents $D_u^{\, \gamma}$ and $D_g^{\, \gamma}$
from both groups at two scales relevant for our comparisons to data.
In both \cite{aurf} and \cite{grvf}, the fragmentation functions are
assumed to evolve from a pure vector meson dominance (VMD) input at some
 very low scale. However, this boundary condition for $D_i^{\, \gamma}
(z,Q^2)$ has been implemented in rather different factorization schemes.
ACFGP use the $\overline{\mbox{MS}}$ scheme, whereas GRV impose the VMD
input in the timelike version of the so-called $\mbox{DIS}_{\gamma}$
scheme, originally introduced for the (spacelike) parton structure of
the photon \cite{grvphot}. In the $\overline{\mbox{MS}}$ scheme employed
here the latter ansatz corresponds to an additional, rather large input
for $D_{q}^{\, \gamma}$, which guarantees the positivity of the timelike
structure function $f_1^{(T)}$ for single photon inclusive $e^+ e^-$
annihilation, $e^+ e^- \rightarrow \gamma X$. Thus the quark-to-photon
fragmentation functions of GRV \cite{grvf} are larger than the ones of
ACFGP \cite{aurf}, especially at low scales, despite the fact that in
\cite{aurf} a sizeably larger VMD input is employed. On the other hand,
$D_g^{\, \gamma}$ of ACFGP is much larger than its GRV counterpart. This
is due to the {\em huge} VMD gluonic input in \cite{aurf}, based on the
assumption $D_g^{\, \rho^{0}}=D_g^{\, \pi^0}$ with $\int_0^1 dz \, z
D_g^{\, \pi^0}(z,2 \mbox{ GeV}^2)=0.5$, meaning that as much as half of
an outgoing gluon's momentum is carried away by neutral pions alone.

The effects of these differences are displayed in Fig.\ 3, where we
show the (scale and scheme dependent) relative importance of the
fragmentation part of the prompt photon cross sections for ISR and
Tevatron energies both for the total and for the gluon initiated
contributions. We have normalized all results to the direct cross
section, and apart from the fragmentation functions, all parameters
and distributions have been chosen as in the `standard' calculation
above.  As already obvious from Fig.\ 1, the ACFGP \cite{aurf} and GRV
\cite{grvf} fragmentation functions yield very similar results for the
total fragmentation contribution. For the isolated Tevatron case,
$g\rightarrow \gamma$ fragmentation plays an almost negligible role due
to the high $z_{min}$ cut implied by the isolation criterion, e.g,
$z_{min}=(1+2 \mbox{ GeV} /p_T)^{-1}$ for CDF conditions. The
$q\rightarrow \gamma$ pieces are rather similar, since $D_q^{\, \gamma}
(z,Q^2)$ is probed at large $z$ and $Q^2$ here. In the fully inclusive
ISR case, on the other hand, $z_{min}=x_T= 2 p_T/\sqrt{s} \stackrel{>}
{{\scriptstyle \sim}}0.15$ for $\eta=0$, and the huge difference in
$D_g^{\, \gamma}$ between \cite{aurf} and \cite{grvf} enters the
fragmentation cross section. However, since the scales $Q^2 \approx
p_T^2/4$ are rather low here, this effect is strongly compensated by
the difference in the quark contributions. Of course, the difference
in the total results is not necessarily fully representative of the
uncertainty originating in the fragmentation part of the cross section.
We have checked that in the theoretically more realistic GRV \cite{grvf}
case, a $50\%$ change in the VMD input distributions for
$D_i^{\, \gamma}$ has no sizeable effect on the results due to the
dominance of $D_q^{\, \gamma}$.  Hence we will keep the fragmentation
functions of \cite{grvf} for the rest of this paper. Clearly,
experimental information on the fragmentation functions, e.g. from
$e^+ e^- \rightarrow \gamma X$ \cite{thom}, is needed.

Let us now discuss the scale dependence of the results, i.e.\ the
changes in the theoretical predictions for varying $\mu_R$ and $\mu_F$.
It turns out that the NLO cross section depends only {\em very} weakly
on $M_F$ both for the isolated \cite{gv1} and the fully inclusive cases.
For the latter, e.g., the difference of the results for $M_F=p_T/2$ and
$M_F=p_T$ at fixed $\mu_R$, $\mu_F$ does not exceed $1\%$. We therefore
keep the fragmentation scale $M_F$ fixed at $M_F=p_T/2$ for the
fixed-target and ISR experiments and $M_F=Rp_T$ for the isolated cross
sections in the following. The dashed and the dash-dotted lines in Fig.\
4 display the shifts in the theoretical results if we choose
$\mu_R=\mu_F=0.3\, p_T$ or $\mu_R=\mu_F=1.0\, p_T$, respectively,
instead of $\mu_R=\mu_F=p_T/2$. More precisely, the curves show
($\sigma_{th}'-\sigma_{th})/\sigma_{th}$, where $\sigma_{th}'$ is the
theoretical cross section as calculated with the new values for the
scales, whereas $\sigma_{th}$ corresponds to the `standard' calculation.
It becomes obvious that the results for $\mu_R=\mu_F=0.3\, p_T$ or
$1.0\, p_T$ amount to almost a constant shift in the normalization of
the theoretical cross section as far as the CDF and UA2 data are
concerned, and do not provide a change in the slope of the cross
section. It can also be seen that the theoretical cross section at
lower energies shows a rather strong scale dependence.  So far, our
results are in agreement with the claims in \cite{cdf,cteqdg} that scale
uncertainties provide (almost) no effect on the $p_T$-shape in the
collider case.

There is, however, no argument that enforces $\mu_R$ and $\mu_F$ to be
exactly equal, they are just expected to be of the same order of
magnitude, given by the prompt photon's $p_T$.  In fact, $\mu_R \neq
\mu_F $ automatically happens if one uses `optimized' scales
\cite{optim} as mentioned in the introduction. A smaller
renormalization scale $\mu_R$ along with a larger factorization scale
$\mu_F$ can be expected to create a steeper slope of the theoretical
result, since lowering $\mu_R$ mainly increases the strong coupling
constant $\alpha_s$, whereas the main effect of a larger $\mu_F$ is to
deplete the gluon distribution at larger $x$ and to increase it at
smaller $x$. In fact, the curves in Fig.\ 4 show that these effects are
quite significant. The choices of, e.g., $\mu_R=0.3\, p_T$, $\mu_F=p_T$
or $\mu_R=p_T$, $\mu_F=0.3 \, p_T$ do lead to about $\pm 20\%$ shape
changes in the CDF region, respectively.  Obviously, scale dependences
{\em are} able to affect the $p_T$-{\em slope} of the NLO cross section,
contradicting the corresponding conclusions drawn in \cite{cdf,cteqdg}
which were derived assuming $\mu_R=\mu_F$. All in all, scale changes
seem to have a rather strong influence on the theoretical cross section
even beyond the LO. This is in line with the observations in \cite{bq},
where it was shown that at small $x_T$ the scale dependence is only
slightly reduced when going from LO to NLO, which renders it difficult
to estimate the most appropriate scales. The scale dependence of the
NLO cross section for prompt photon production indicates the importance
of corrections of even higher order and sets severe limits on the
accuracy of gluon determinations from these data.
\section{Combined Analysis of DIS and Prompt Photon Data}
In this section, we examine the question of whether the agreement
between NLO calculation and the isolated prompt photon data can be
further improved by adapting the proton's parton content, in particular
its gluon density. For this purpose, we perform combined NLO fits to
direct-$\gamma$ production and DIS structure function data, the
latter pinning down the quark densities, with different choices for
the renormalization and factorization scales. In this way, we also
investigate the uncertainty of the resulting gluon distribution
originating from the scale dependence of the prompt photon cross
section.

Technically we proceed as follows: at the reference scale of $Q_0^2=4$
GeV$^2$ the gluon input is parametrized as
\begin{equation}
xg(x,Q_0^2)=A_g x^{\alpha_g} (1-x)^{\beta_g}
\left( 1+ \gamma_g \sqrt{x}+\delta_g x \right) \:\:\: .
\end{equation}
This functional form is also used in the latest MRS analysis \cite{mrs}.
For each given set of $\alpha_g,\, \beta_g,\, \gamma_g$, and $\delta_g$,
a fit of the quark densities to $F_2^{\, p}$ data is performed, where
$A_g$ is fixed by the energy-momentum sum rule. Here the non-singlet
quark densities $u_v=u-\bar{u}$, $d_v=d-\bar{d}$ and $\Delta =\bar{d}-
\bar{u}$ are, as in \cite{grv}, for $\Lambda_{\overline{\rm MS}}^{(4)}
=200 $ MeV directly adopted from the MRS(A) global fit \cite{mrs}. For
other values of $\Lambda$ we employ the modified non-singlet sets of
\cite{vogt}. This procedure guarantees a sufficiently accurate
description of all observables testing mainly the flavor decomposition
of the quark content and facilitates the fitting procedure since it
reduces the number of free parameters. The sea quark input is
parametrized as \cite{mrs}
\begin{eqnarray}
x(\bar{u}+\bar{d})(x,Q_0^2)&=&A_{\xi} x^{\alpha_{\xi}}
(1-x)^{\beta_{\xi}} \left( 1+\gamma_{\xi} \sqrt{x}+
\delta_{\xi} x \right) \:\:\: , \\
xs(x,Q_0^2) &=& \frac{1}{4}x(\bar{u}+\bar{d})(x,Q_0^2) \nonumber\:\:\: .
\end{eqnarray}
This input is evolved in the factorization scheme of \cite{grv} (see
section 3), calculating the charm contribution to $F_2$ via the LO
Bethe-Heitler process using the LO gluon distribution and $\Lambda_{LO}
$ of \cite{grv}.
The five parameters in eq.\ (4) are fitted to the available $F_2^{\, p}$
data of BCDMS \cite{bcdms,milst}, NMC \cite{nmc}, ZEUS \cite{zeusf2},
and H1 \cite{h1f2} in the region where the structure function is
sensitive to the sea quark and gluon densities, $x \leq 0.3$, and where
higher twist contributions are expected to be small, $ Q^2\geq 5 $
GeV$^2$.  As in \cite{mrs} the BCDMS data are normalized down by
$ 2\% $. At the present level of experimental accuracy, the HERA
$F_2^{\, p}$ normalization uncertainties \cite{zeusf2,h1f2} can be
disregarded for our purpose. Statistical and systematic errors are
added quadratically for all data sets.

We then use the complete set of parton distributions obtained from this
$F_2^{\, p}$ fit for a fixed gluon shape to determine the $\chi_{dir.\,
\gamma}^2$ for the prompt photon data of WA70, R806/7, UA2 and CDF
\cite{wa70,r807,ua2,cdf}, already discussed in section 3, supplemented
by the $pp$ inclusive results of NA24 \cite{na24} and UA6 \cite{ua6},
both at $\sqrt{s} = 24 $ GeV.
Due to the very small point-to-point errors of the CDF data \cite{cdf}
their overall normalization uncertainty is rather important. Therefore
it is allowed to float in the fit with a contribution to $\chi_{dir.\,
\gamma}^2$ according to its experimental uncertainty of $ 10\% $, see
\cite{cteq}. All other data sets, including \cite{na24,ua2} where
separate normalization uncertainties are provided, are fixed at their
nominal normalization. The whole procedure is repeated until finally
$\chi_{tot}^2 = \chi_{DIS}^2 +\chi_{dir.\, \gamma}^2$ is minimized. We
have chosen this two-step approach of two `nested' fits since it affords
the least number of evaluations of $\chi_{dir.\, \gamma}^2$ which
dominate the consumed computer time due to the complexity of the NLO
prompt photon cross section calculation.
In order to examine the uncertainty coming from the choice of the QCD
scale parameter $\Lambda$, we perform fits for $\Lambda_{\overline{\rm
MS}}^{(4)}=200$ MeV and 300 MeV. These values are representative of the
present range of $\Lambda$ found in analyses of DIS and related data
\cite{mrs,milst,nmcqcd,ccfr}. In view of our findings for the scale
dependence of the prompt photon cross section obtained in the last
section, we repeat the fits for various combinations of the
renormalization/factorization scales, systematically scanning the range
$ 0.3\, p_T\leq \mu_R, \mu_F \leq 2.0\, p_T $, see below for a detailed
discussion.

Fig.\ 5 displays the results of three of our fits, again compared to
the data of \cite{wa70,r807,ua2,cdf}. For clarity, the additional
large-$x_T$ data sets of \cite{na24,ua6} are not shown in the figure.
The upper part presents the fit using the `standard' scales and
$\Lambda_{\overline{\rm MS}}^{(4)} $ of section 3, the lower plot
depicts further examples of best possible fits, one for each
$\Lambda $-value employed. The latter two fits happen to lead to the
same CDF normalization factor of $+10\% $. The input parameters and
$\chi^2 $-results for these three representative fits are given in
Table 1. It is obvious that we obtain good fits, with $\chi^2 $ per
data point slightly below 1 also in the prompt photon subset of the
fitted data. The ISR and fixed-target data \cite{wa70,na24,r807,ua6}
are described about as well as in the reference calculation in section
3. The $\chi^2 $ for the CDF results \cite{cdf} seems to remain a bit
high. We have not been able to reach a value below 25 for the 16 data
points, however with more than 10 units contributed by just the
measurement at $ p_T = 48.9 $ GeV. The figure clearly demonstrates that
{\em no} shape problem remains here. With respect to the UA2 results
\cite{ua2}, the situation does only slightly improve; with present
errors, however, the UA2 data do not impose strong constraints on the
fits.
On the other hand, it should be noted that a data set with about the
same central values but much smaller errors would, in combination with
the CDF results, very seriously challenge the NLO framework used in this
paper, since we have been unable to improve the description sizeably
even by artificially reducing the UA2 errors in special runs of our
fits.
For the larger value of $\Lambda_{\overline{\rm MS}}^{(4)}$, the range
of renormalization and factorization scales giving very good fits to the
data is somewhat shifted towards larger values. As can be seen from
Fig.\ 5 and the table, we do not find a significant difference in the
quality of the fits for the two $\Lambda$-values. Taking into account
the full scale dependence of the NLO cross section reduces the
sensitivity of the prompt photon data to $\alpha_S $ as compared to,
e.g., \cite{aur1,mrsl} where `optimized' scales were used.

\begin{table}[htb]
  \begin{center}
  \begin{tabular}{|c||r|r|r|}\hline
                              &        &        &        \\[-0.4cm]
  $\Lambda^{(4)}/ \mbox{MeV} $&   200  &   200  &  300   \\[0.1cm]\hline
                              &        &        &        \\[-0.4cm]
  $ \mu_{R} / p_T $           &   0.5  &   0.3  &  0.7   \\
  $ \mu_{F} / p_T $           &   0.5  &   0.7  &  0.7   \\[0.1cm]\hline
                                                                  \hline
                              &        &        &        \\[-0.4cm]
  $ \alpha_{g} $              & -0.011 & -0.125 & -0.080 \\
  $ \beta_{g}  $              &  5.715 &  5.314 &  5.024 \\
  $ \gamma_{g} $              & -4.174 & -4.200 & -3.963 \\
  $ \delta_{g} $              &  5.217 &  5.370 &  4.931 \\[0.1cm]\hline
                              &        &        &        \\[-0.4cm]
  $ A_{\xi} $                 &  0.768 &  0.775 &  0.787 \\
  $ \alpha_{\xi} $            & -0.156 & -0.148 & -0.135 \\
  $ \beta_{\xi}  $            &  7.480 &  7.641 &  7.173 \\
  $ \gamma_{\xi} $            &  0.642 &  0.740 &  0.843 \\
  $ \delta_{\xi} $            &  1.521 &  1.622 &  1.080 \\[0.1cm]\hline
                                                                  \hline
                              &        &        &        \\[-0.4cm]
  $ \chi^2_{DIS} $            &  253.6 &  257.3 &  254.4 \\
  $ \chi^2_{dir.\, \gamma} $  &   59.1 &   52.8 &   56.5 \\
  $ \chi^2_{tot}$             &  312.7 &  310.1 &  310.9 \\[0.1cm]\hline
  \end{tabular}
  \caption{Input parameters for eqs.\ (3) and (4) for three
    representative combined fits to direct-$\gamma $ and
    $ F_{2}^{\, p} $ data, together with the resulting $ \chi^2
    $-values for the 60 prompt photon and 294 DIS data points.}
  \end{center}
\end{table}

Our results demonstrate that presently published data on $pp,\, p\bar{p}
\rightarrow \gamma X $ {\em can} be described quantitatively by NLO
perturbative QCD. This finding is at variance with the results of a
recent, partly comparable analysis of the CTEQ group \cite{cteqdg}.
In that paper, it is concluded that (at least) the $p_T$-shapes of the
ISR and collider data cannot be satisfactorily fitted, unless an
`intrinsic' $k_T$-smearing is introduced, which is as large as $<\! k_T
 \! > \, \approx 4 $ GeV for $Sp\bar{p}S$ and Tevatron conditions.
It should be clear that we do not claim positive evidence for the
absence of such a (somewhat counterintuitive) smearing, we just state
that the data considered here can be accounted for without this
assumption. It should be noted in this context, however, that a
$ <\! k_T \! > $ of about 1 GeV for fixed-target energies would spoil
\cite{stir} the successful description of the WA70 data \cite{wa70}.
The consequences of our optimal `pure' NLO fits to all direct photon
data will be elucidated below.

Let us note before some obvious differences between our study and \cite
{cteqdg}. With respect to data, we use the latest results of R806/7
\cite{r807}, whereas \cite{cteqdg} includes previous superseded steeper
results of this collaboration \cite{r806}. Taking into account the
results of \cite{r807}, there is obviously no slope offset between NLO
theory and ISR data, neither for `standard' distributions, see Fig.\ 1
and the last reference of \cite{mrs}, nor for our fits discussed above.
In the fixed-target region, we do not include the recent E706 data
\cite{e706} from {\em pBe} collisions in order to avoid any bias
from a possible EMC-like effect on the nuclear gluon density.
With respect to the theoretical treatment, in contrast to \cite{cteqdg}
we fully incorporate {\em NLO} fragmentation and the corresponding
treatment of isolation into our analysis. The very small errors of the
CDF data \cite{cdf} make a complete and consistent NLO analysis
mandatory in order to arrive at any solid statements on the viability
of the NLO description of prompt photon production.\footnote{The input
form used in the fits of \cite{cteqdg} has not been given in the paper.
In the likely case that the CTEQ3 \cite{cteq} input ansatz has also been
employed in \cite{cteqdg}, another difference with respect to our study
would be added, since we find that the inclusion of the $\sqrt{x}$-term
in (3), not present in \cite{cteq}, is important.}

Hence in the present situation it seems reasonable to us to maintain
the `pure' NLO perturbative QCD framework and to investigate the
consequences of optimal fits to the direct-$\gamma $ data in this
scenario. In Fig.\ 6 the gluon distributions obtained in the three fits
displayed above are compared at $ Q^2 = 20 \mbox{ GeV}^2 $ to the
corresponding result of the recent MRS(A$^{\prime}$) global study
\cite{mrs} where only fixed-target prompt-$\gamma$ data were included
in the analysis.
Although data at different $p_T$ probe the parton distributions at
different scales, the main effect of including the CDF results \cite
{cdf} is readily seen in the figure. The gluon distribution is increased
at the lower end of the $x$-range accessed by these measurements, $ x
\simeq 0.01 $, and is decreased around $ x \simeq 0.15 $, thereby
in between providing the steeper rise suggested by the data. At very
small $x$, $ x \stackrel{<}{{\scriptstyle \sim}} 10^{-3} $, our gluon
densities are larger than the one of MRS(A$^{\prime}$) and more similar
in magnitude to the MRS(G) \cite{mrs} and GRV \cite{grv}
parametrizations.

Effects of such an enhanced gluon density at $x \approx 10^{-2}$ should
also show up in other processes which are sensitive to $ g(x,Q^2) $
in that $x$-region such as, e.g, bottom production in high-energy $p
\bar{p} $ collisions at the Tevatron. Fig.\ 7 illustrates this fact.
The cross section for single-inclusive $b$-quark production, $\sigma(p
\bar{p}\rightarrow bX) $, expected to be mainly driven by gluon-gluon
fusion, has been measured as a function of the minimal transverse
momentum $p_T^{\, min}$ of the $b$-quark by the CDF and D0
collaborations recently \cite{cdfbo,d0bo}. The figure shows these data
together with corresponding NLO calculations based on the work of
\cite{dawson,laenen} for two of our fitted parton distribution sets and
the MRS(A$^{\prime})$ parametrization \cite{mrs}\footnote{We thank M.
Stratmann for performing these calculations.}. For this illustration,
we have chosen $ \mu = \frac {1}{2} \sqrt{p_T^2 + m_b^2} $ for the
renormalization and factorization scales and $ m_{b} = 4.5 $ GeV for
the bottom mass. Note that the theoretical uncertainties from variations
of these parameters are rather substantial \cite{cdfbo,d0bo}.

Let us finally come back to the scale dependence of the fitted gluon
densities. As already mentioned above, we have scanned the scale
range $ 0.3 \, p_T \leq \mu_{R},\, \mu_{F} \leq 2.0 \, p_T $ in our
fits. The lower limit has been introduced in order to avoid low scales
below about 1 GeV. We find very good combined fits to the complete set
of prompt photon production and DIS structure function data given above
for $ 0.3 \, (0.5) \stackrel{<}{{\scriptstyle \sim}} \mu_{R},\, \mu_{F}
\, / p_T \stackrel{<}{{\scriptstyle \sim}} 0.7 \, (1.0) $ for $ \Lambda
^{(4)} = $ 200 MeV (300 MeV), respectively, and $ \mu_{R} \stackrel{<}
{{\scriptstyle \sim}} \mu_{F} $. The spread of the gluon distributions
of all these best fits can serve as a measure for the theoretical
uncertainty on $ xg(x,Q^2) $ induced by scale variations. Fig.\ 8
displays the resulting `error band', normalized to the MRS(A$^{\prime
}$) gluon distribution \cite{mrs} at $ Q^2 = 20 \mbox{ GeV}^2 $. Also
shown are the uncertainties at four selected $x$-values arising from
the experimental errors.
They have been determined from fits for one scale combination, which
are one unit in the total $ \chi^2 $ higher than the corresponding best
fit. In view of the controversial theoretical interpretation \cite
{ggrv,cteqdg} of the isolated high-energy collider data \cite{ua2,cdf},
we have repeated the fits for the theoretical error band retaining only
the 30 (non-isolated) fixed-target and ISR data points of
\cite{wa70,na24,r806,ua6} in the direct-$\gamma $ part of the analyses.
The resulting band is also presented in Fig.\ 8.

It becomes obvious from the figure that the same pattern of deviations
from the MRS(A$^{\prime}$) fit \cite{mrs}, namely an increase (decrease)
at $x$ around 0.01 (0.15), is present for all scale choices which allow
for a good fit of the complete set of prompt photon data. Thus this
effect is significant and cannot be removed by a `better' choice of
scales.
Not too surprisingly, the fits restricted to the non-isolated data
lead to gluon densities considerably closer to standard distributions.
They are, taking into account the experimental uncertainties, in fact
in good agreement with, e.g., the GRV gluon parametrization \cite{grv}
employed as a standard choice in section 3. At $ x < 0.1 $, the
experimental uncertainties and the theoretical error bands of the fits
to all data are of about the same size. However, in the classical
fixed-target regime, $ x > 0.3 $, mainly considered in \cite{aur1} and
many subsequent analyses via `optimized' scales \cite{optim}, the
uncertainty of $g(x,Q^2)$ is dominated by scale variations. For
instance, at $ x \simeq 0.4 $, the spread of the fitted gluon densities
is as large as almost a factor of two. In this large-$x$ region, shown
separately in Fig.\ 9 also for a high scale typical for the production
of particles with masses of a few hundred GeV, a comparison of recent
parametrizations, e.g.\ of MRS(A$^{\prime}$,G) and GRV \cite{grv} as
displayed in Fig.\ 8, does not indicate the real uncertainty of the
gluon distribution.
\section{Conclusions}
We have performed a detailed NLO perturbative QCD study of prompt
photon production in $pp$ and $p\bar{p}$ collisions with respect to
its sensitivity to the proton's gluon content. In fact, our analysis
is the first one that includes both the full NLO treatment of the
fragmentation contribution and the complete NLO treatment of isolation,
and the consideration of the full information available now from
experiment.
We have carefully studied the theoretical uncertainties, especially
those arising from the so far experimentally unknown parton-to-photon
fragmentation functions and from scale variations.

The dependence of the theoretical cross sections on the renormalization
and factorization scales turns out to be the dominant source of
uncertainty. Particularly at large-$x$, $x > 0.3$, the scale dependence
very severely limits the accuracy of NLO gluon determinations from
(fixed-target) prompt-$\gamma $ production data. Present-day gluon
parametrizations of CTEQ3, GRV, and MRS(A$^{\prime}$,G) are very
similar in this region and their difference does not represent the real
uncertainty of $ g(x,Q^2) $ here. The isolated direct-$\gamma $ cross
sections are also sizeably affected by the choice of these scales in
magnitude and $p_T$-shape.

We have carried out combined global analyses of fixed-target and
collider prompt-photon cross sections and DIS structure function data
within the perturbative NLO framework. We find that such combined fits
work very well and give a good description of all prompt photon data
for various combinations of renormalization and factorization scales
and values of the QCD scale parameter $\Lambda $. The gluon
distributions resulting from such fits turn out to be larger at $ x
\approx 0.01 $ and smaller at $x$ around 0.15 than those obtained in
recent global analyses of parton distributions not taking into account
the full set of prompt photon data.
\section*{Acknowledgements}
We thank M. Gl\"{u}ck and E. Reya for useful discussions.
This work was supported in part by the German Federal Ministry for
Research and Technology under contract No.\ 05 6MU93P.
%
%

%
%
\section*{Figure Captions}
\begin{description}
\item[Fig.\ 1:] The `default quantity' $(\sigma_{exp}-\sigma_{th})/
\sigma_{th}$ vs.\ $x_T=2 p_T/\sqrt{s}$ for the data of
\cite{wa70,r807,ua2,cdf} as compared to the NLO theoretical cross
section $\sigma_{th}$, using the GRV parton distributions and photon
fragmentation functions \cite{grv,grvf}. The curves present the shifts
$(\sigma_{th'}-\sigma_{th})/\sigma_{th}$, where $\sigma_{th'}$ denotes
the theoretical cross section if the fragmentation contribution is
neglected or if the fragmentation functions of \cite{aurf} are used.
\item[Fig.\ 2:] Comparison of the fragmentation functions
$z D_u^{\, \gamma}(z,Q^2)/\alpha$ and $z D_g^{\, \gamma}(z,Q^2)/\alpha$
from ACFGP \cite{aurf} and GRV \cite{grvf} at $Q^2=10$ and 100
$\mbox{GeV}^2$.
\item[Fig.\ 3:] NLO $\overline{\mbox{MS}}$ fragmentation piece for the
 ACFGP \cite{aurf} and GRV \cite{grvf} photon fragmentation functions
 at ISR and Tevatron energies, using the CDF isolation criterion in
 the latter case. Also shown are the contributions stemming from
 gluon-to-photon fragmentation only. All cross sections are normalized
 to the NLO direct piece. The parton distributions of \cite{grv} are
 used and the scales are $\mu_R=\mu_F=0.5\, p_T$.
\item[Fig.\ 4:] Same as Fig.\ 1, but the lines displaying the shifts
in the theoretical results if the renormalization and factorization
scales are varied as indicated in the figure.
\item[Fig.\ 5:] The data of \cite{wa70,r807,ua2,cdf} as compared to
 three NLO fits for the different choices of $\Lambda_{\overline
 {\rm MS}}$ and of the scales indicated in the figure.
 The CDF data \cite {cdf} are shown with their fitted normalizations.
\item[Fig.\ 6:] Comparison of the gluon distributions obtained in the
 three fits displayed in Fig.\ 5 with the MRS(A$^{\prime}$) \cite{mrs}
 gluon density. $ \overline{\Lambda } $ denotes $ \Lambda^{(4)}_
 {\overline{\rm MS}} $ in MeV.
\item[Fig.\ 7:] The cross section for single-inclusive bottom
 production, $\sigma(p \bar{p}\rightarrow bX) $, as a function of the
 minimal transverse momentum $p_T^{\, min}$ of the $b$-quark as
 calculated in NLO perturbative QCD using the parton distributions
 indicated in the figure. Also shown are the (inclusive lepton) data
 of \cite{cdfbo,d0bo}.
\item[Fig.\ 8:] The error bands of the fitted gluon densities due to
 scale variations for the fit to all data and for the fit retaining the
 (non-isolated) fixed-target \cite{wa70,na24,ua6} and ISR \cite{r807}
 prompt photon data only, plotted via the deviation from the
 MRS(A$^{\prime}$) fit. Also displayed are the $1\sigma$ uncertainties
 resulting from the experimental errors and two further `standard'
 gluon parametrizations \cite{mrs,grv}.
\item[Fig.\ 9:] Same as Fig.\ 8, but for the absolute gluon
 distributions at large $x$ at a typical scale probed by the direct-$
 \gamma $ data (left) and at a very large scale relevant for the
 production of very heavy particles (right). Of the rather similar
 `standard' gluons only the one of MRS(A$^{\prime}$) \cite{mrs} is
 shown for comparison.
\end{description}

\newpage
\vspace*{3cm}
\hspace*{-0.75cm}
\epsfig{file=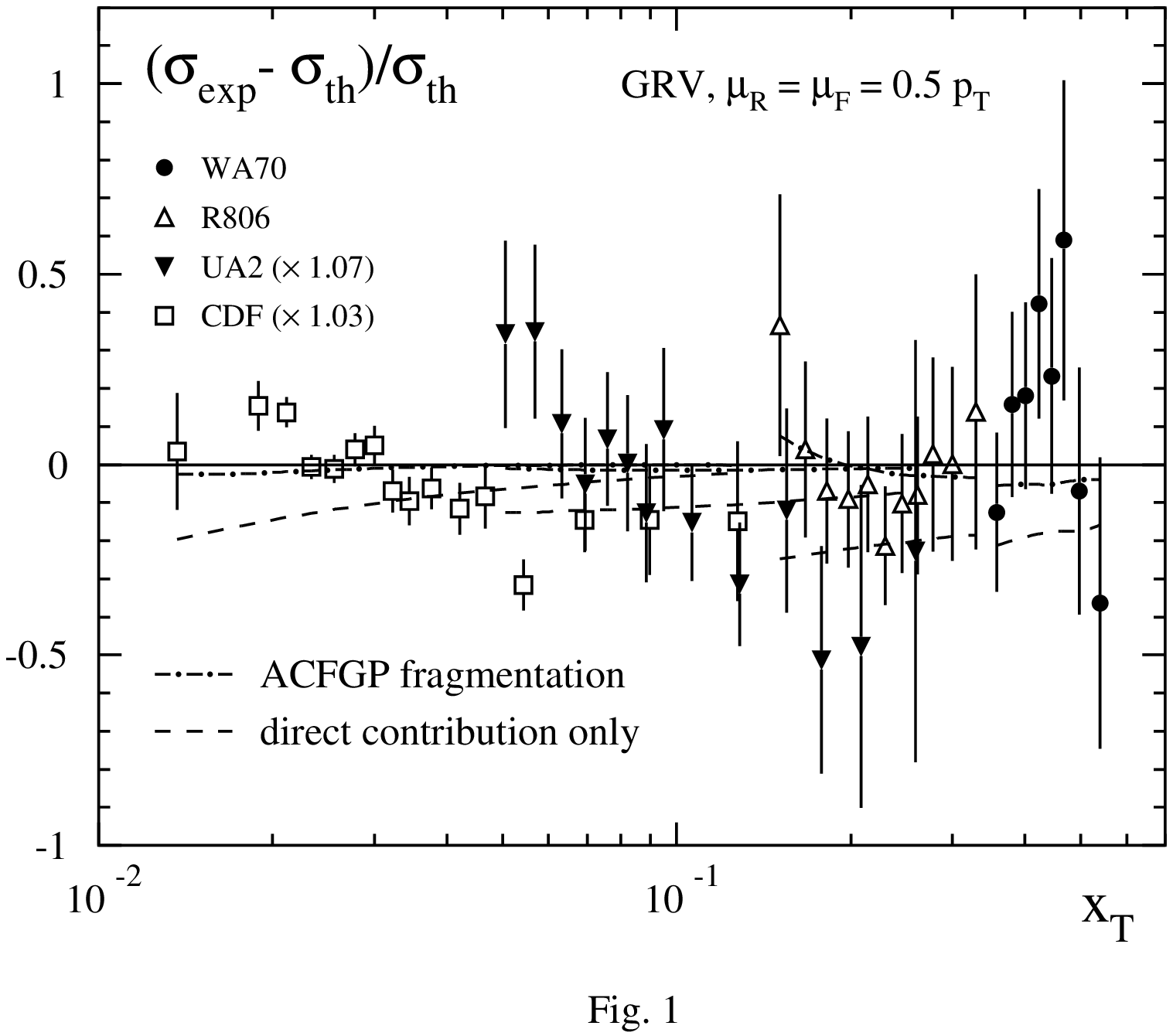,width=16cm}

\newpage
\vspace*{-2.5cm}
\hspace*{-0.3cm}
\epsfig{file=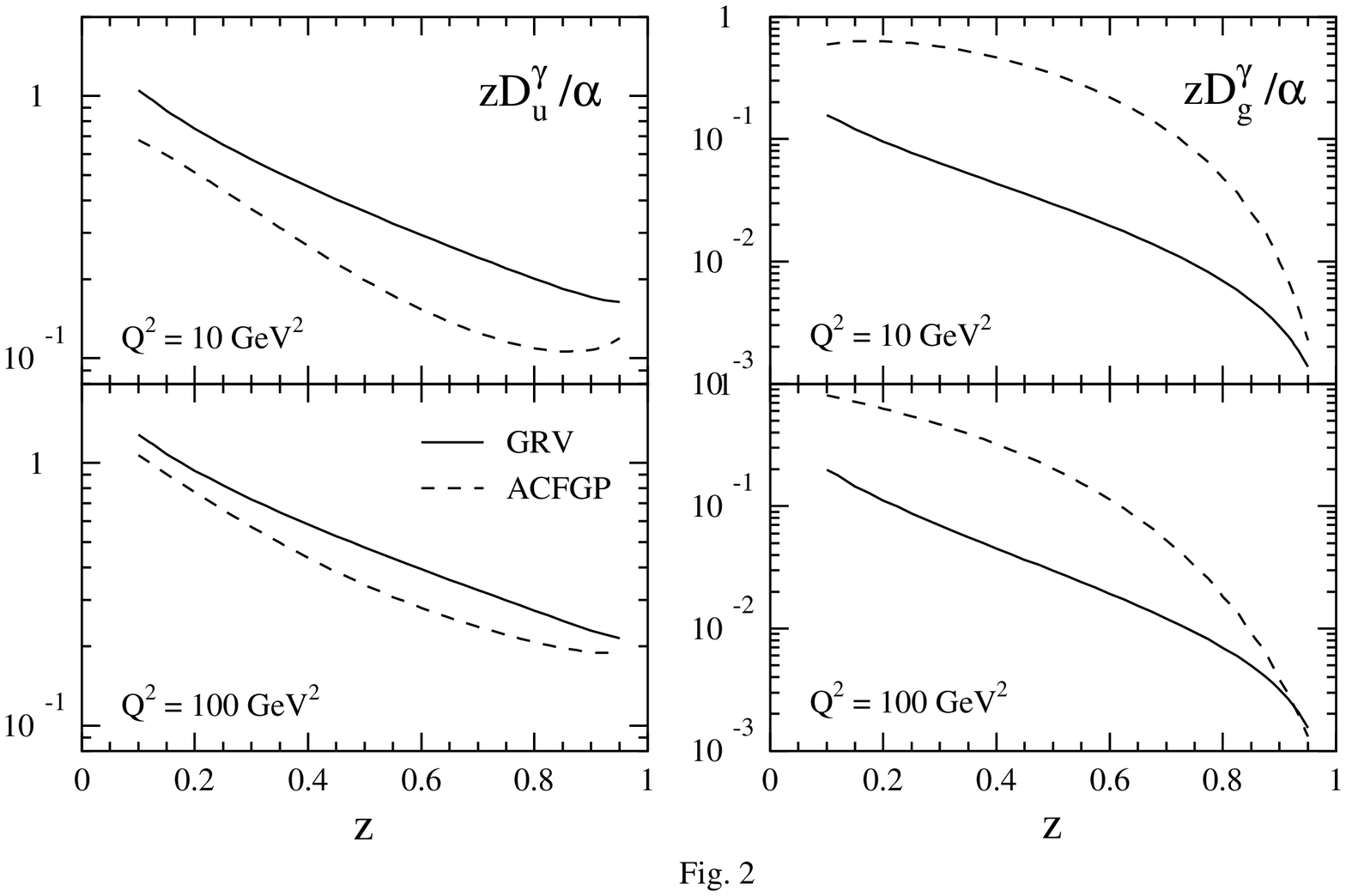,width=15.5cm,angle=90}

\newpage
\vspace*{-2.5cm}
\hspace*{-0.3cm}
\epsfig{file=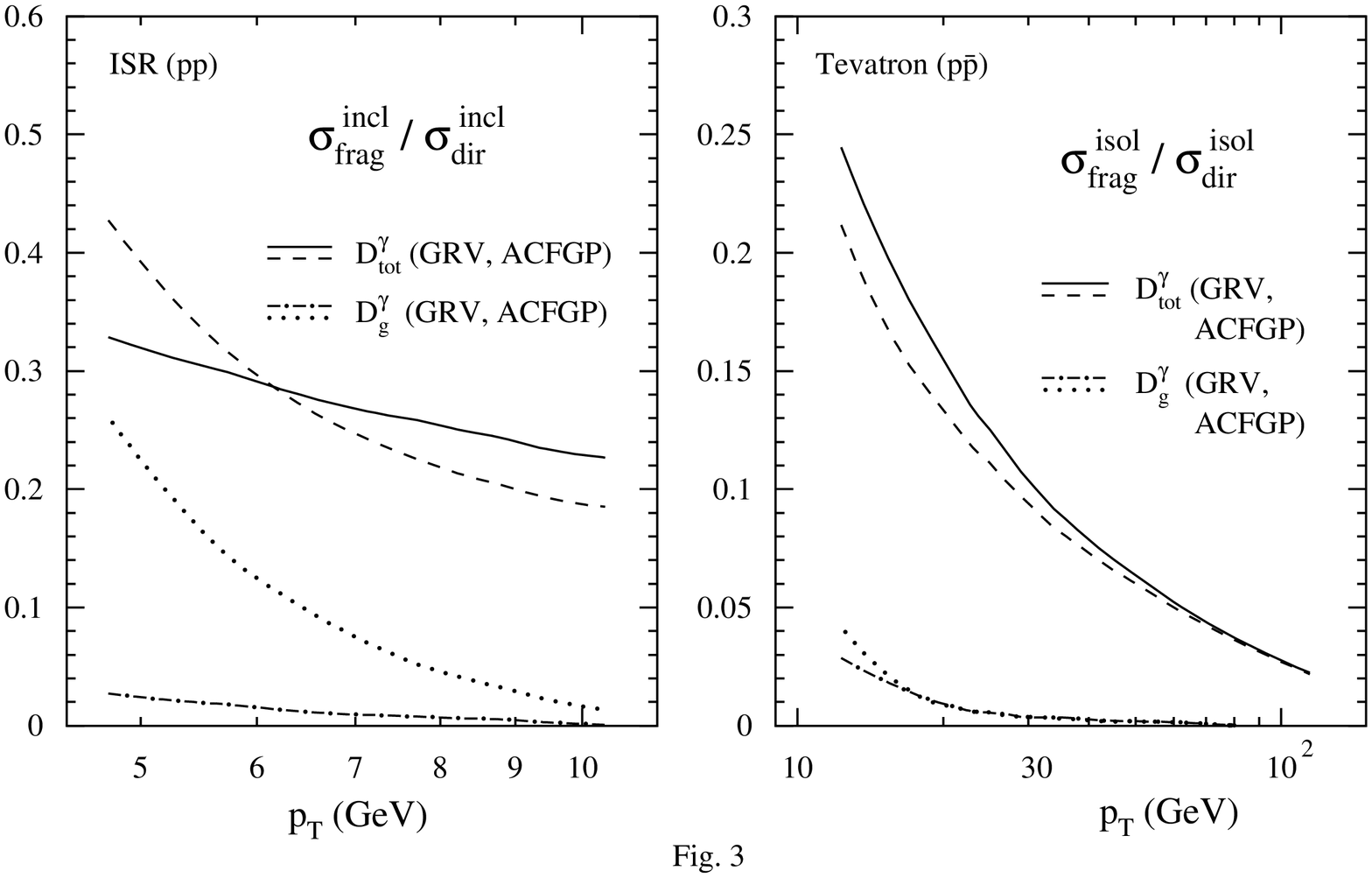,width=15.5cm,angle=90}

\newpage
\vspace*{-3.5cm}
\hspace*{-0.75cm}
\epsfig{file=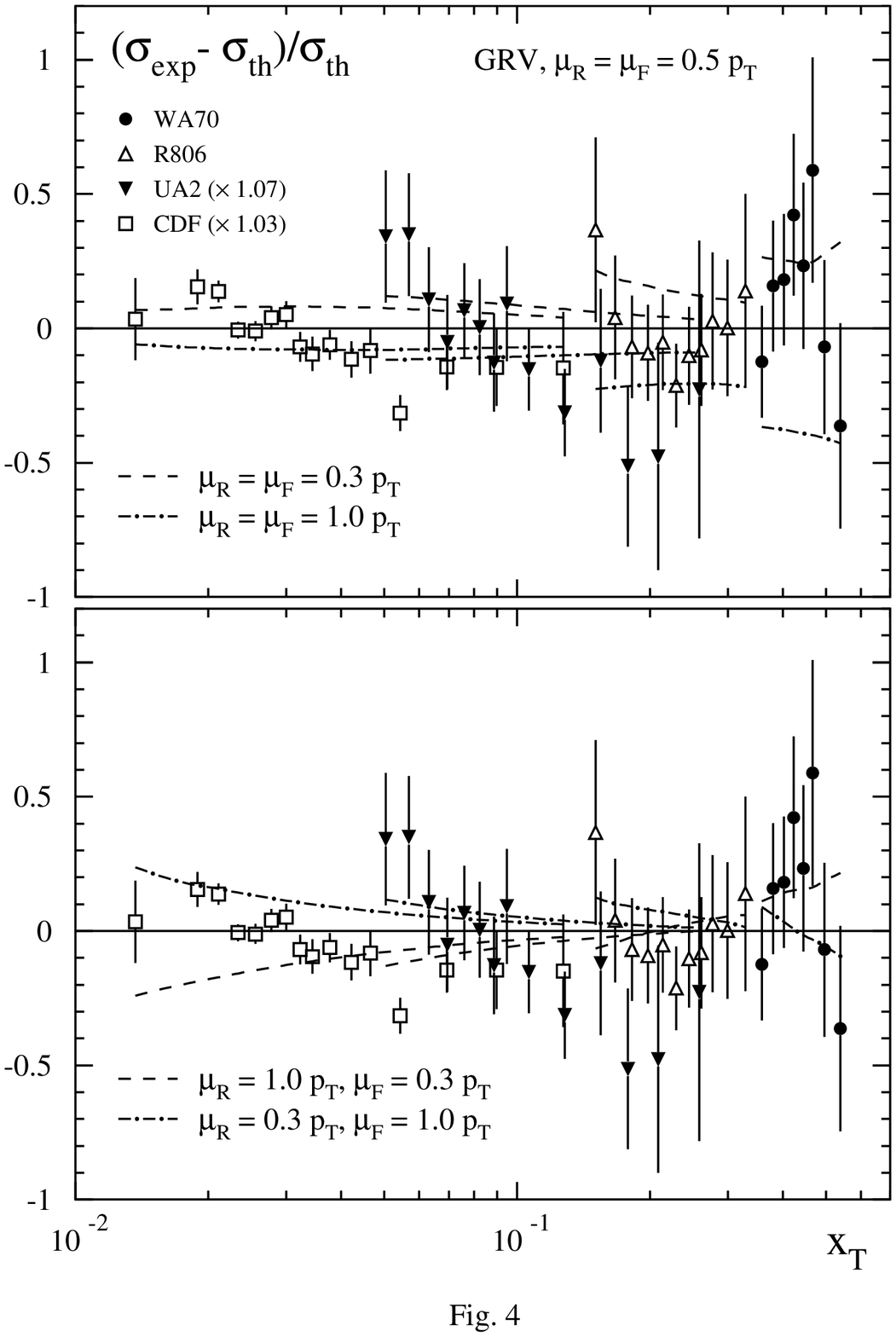,width=16cm}

\newpage
\vspace*{-3.5cm}
\hspace*{-0.75cm}
\epsfig{file=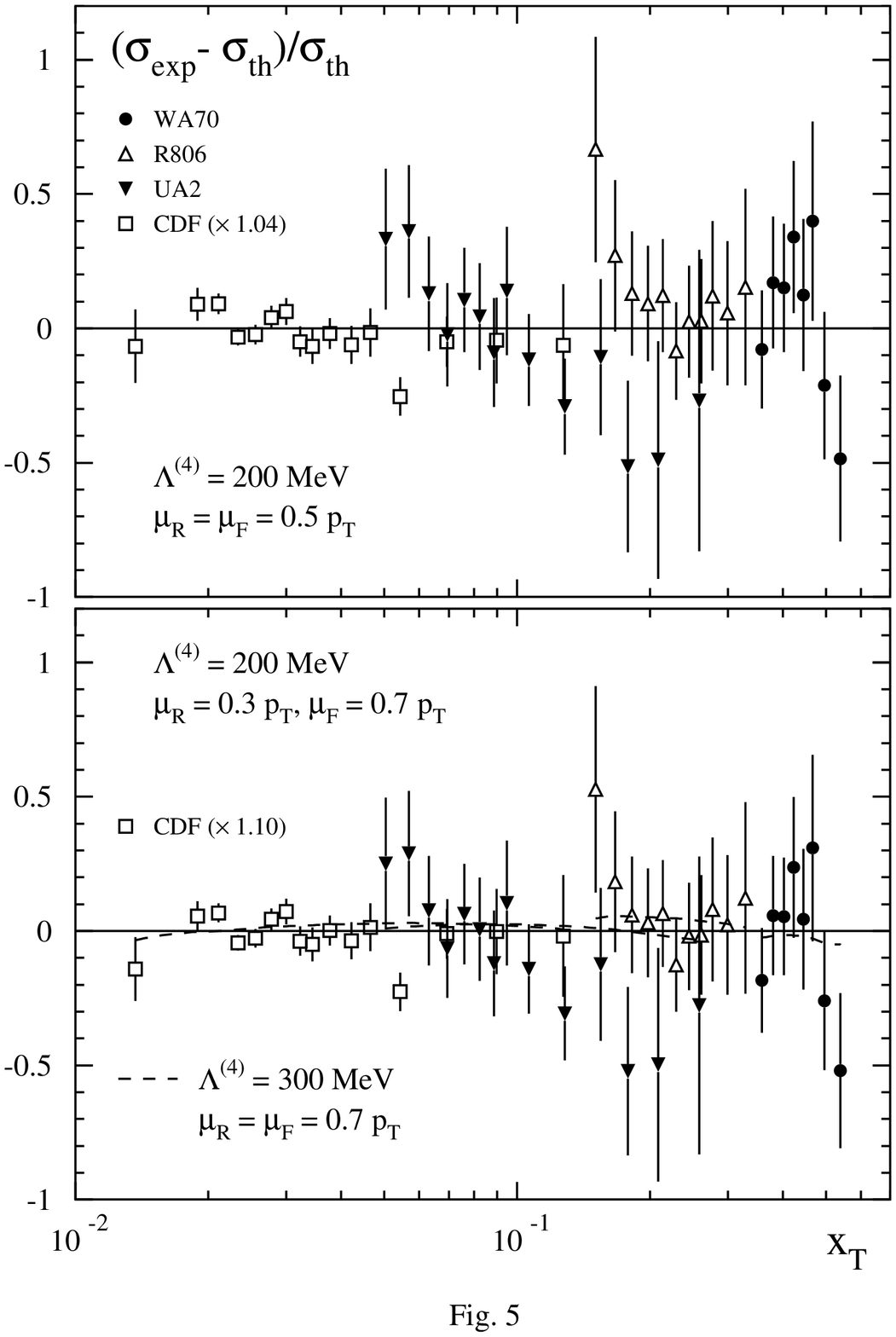,width=16cm}

\newpage
\vspace*{2cm}
\hspace*{0.8cm}
\epsfig{file=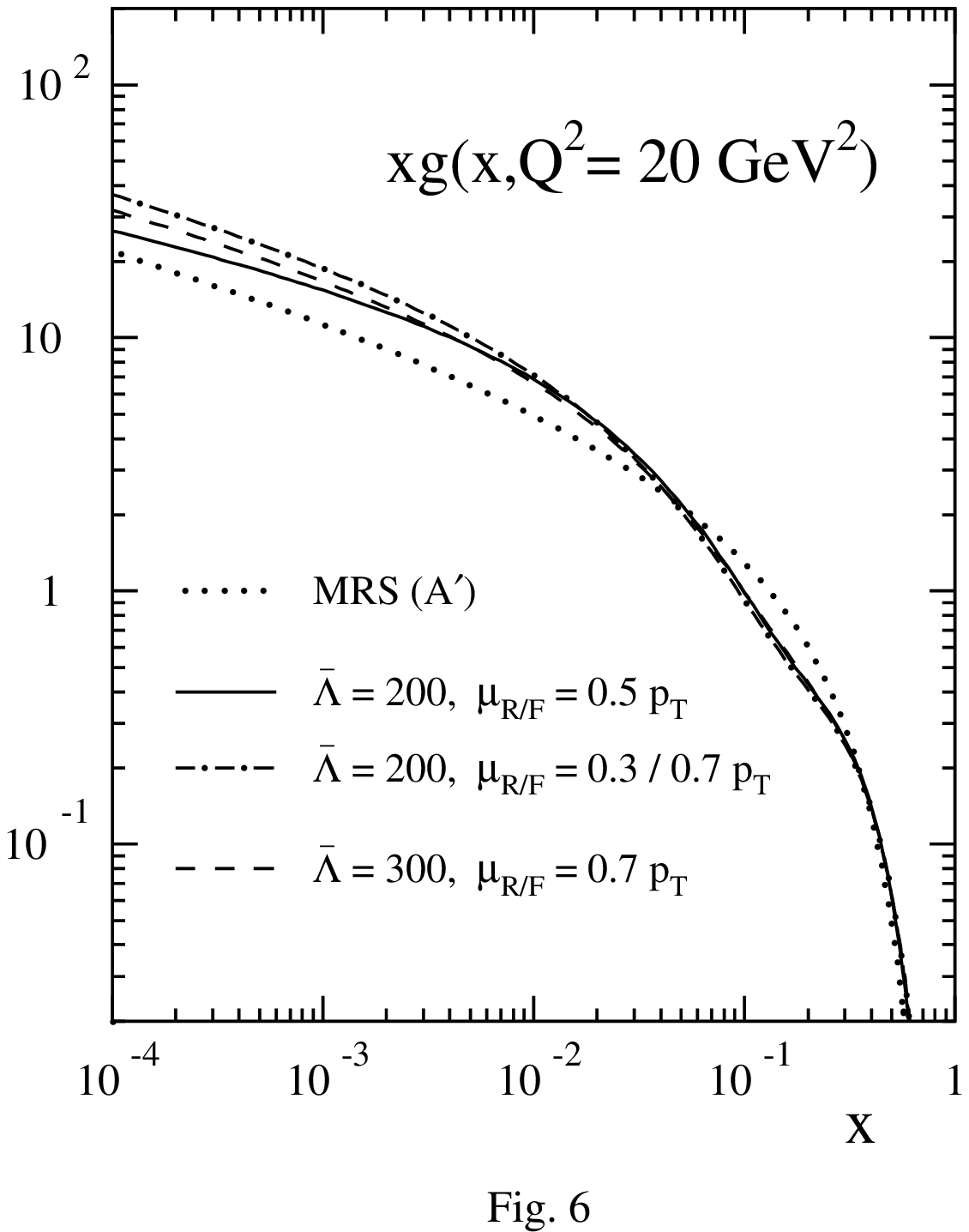,width=13cm}

\newpage
\vspace*{2cm}
\hspace*{0.8cm}
\epsfig{file=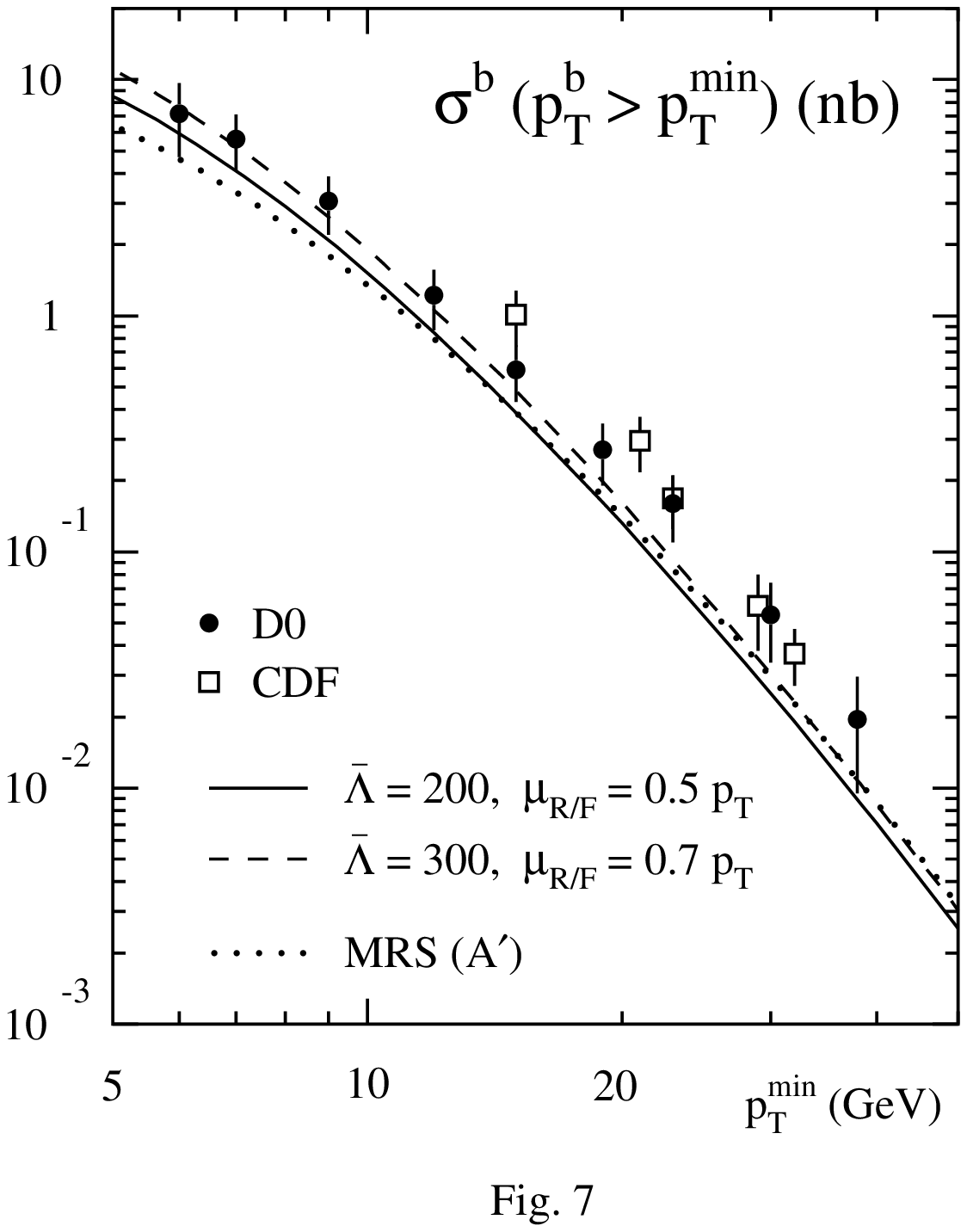,width=13cm}

\newpage
\vspace*{3cm}
\hspace*{-0.75cm}
\epsfig{file=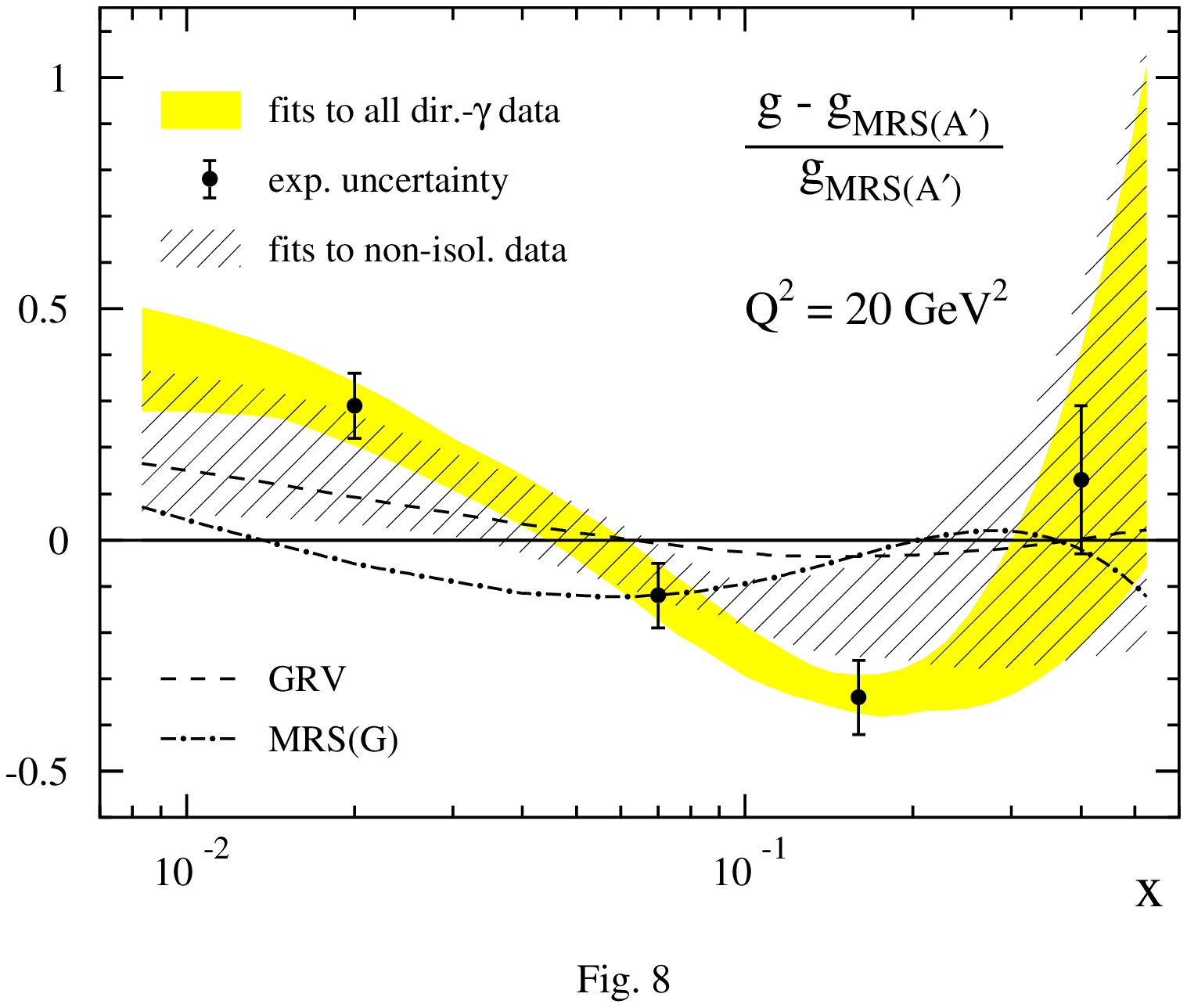,width=16cm}

\newpage
\vspace*{-2.5cm}
\hspace*{-0.3cm}
\epsfig{file=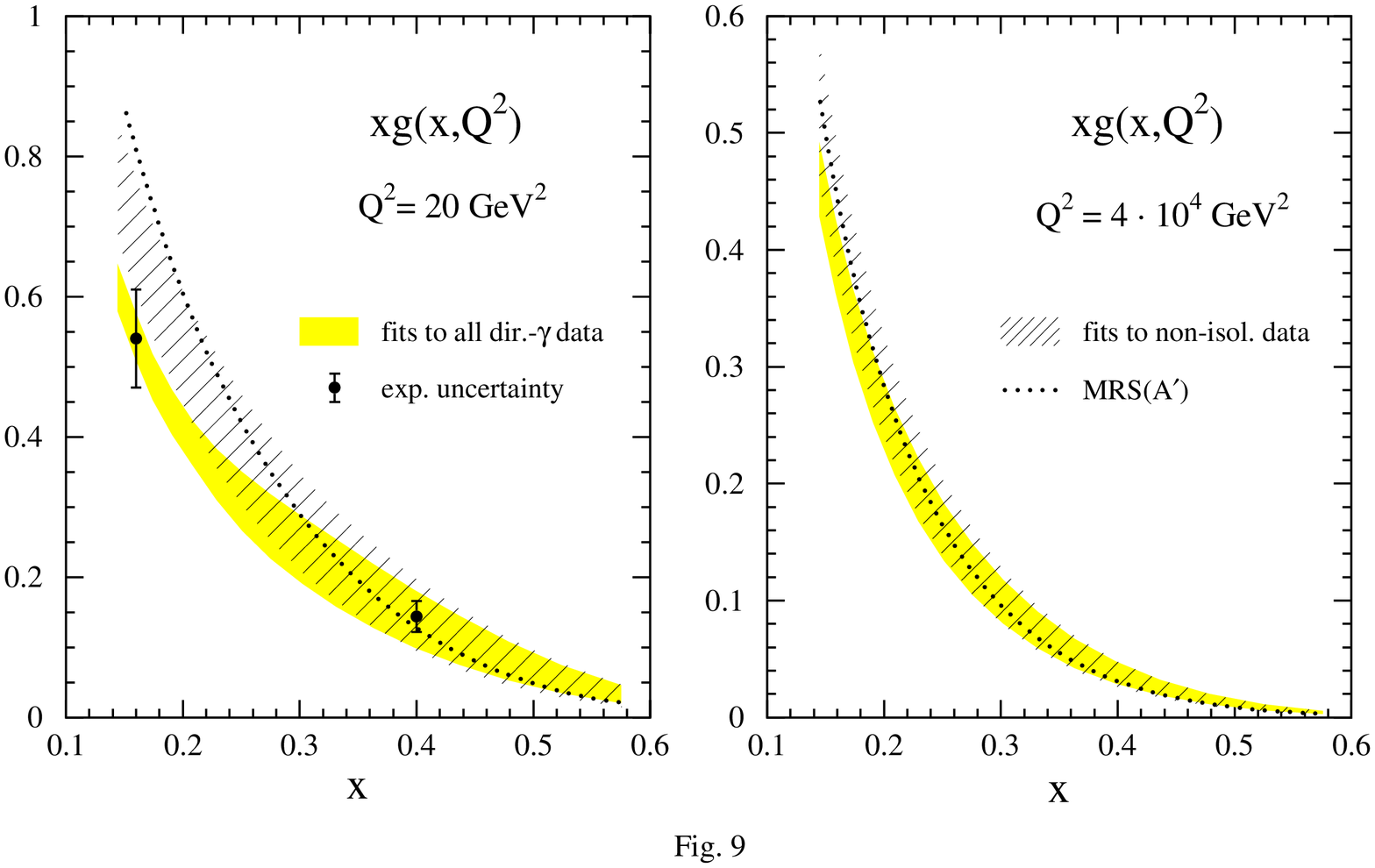,width=15.5cm,angle=90}

\end{document}